\documentstyle[epsfig,twocolumn,prl,aps]{revtex}

\newcommand{\be}{\begin{equation}}
\newcommand{\ee}{\end{equation}}
\newcommand{\bea}{\begin{eqnarray}}
\newcommand{\eea}{\end{eqnarray}}
\renewcommand{\d}{{\rm d}}

\begin{document}                                                
\title{Color Glass Condensate at RHIC~?}
\author{Adrian Dumitru}
\address{
Department of Physics, Brookhaven National Laboratory,
Upton, New York 11973-5000\\
{\small email: dumitru@quark.phy.bnl.gov}\\
}
\date{\today}
\maketitle
\begin{abstract} 
Heavy-ion collisions at the BNL-RHIC collider can probe
whether gluon saturation effects in nuclei at small $x$ have set in, or
whether leading-twist perturbative estimates of particle production are still
applicable. I discuss that soon to come data from $pA$ collisions
at RHIC may provide more systematic insight into this problem than present
data on the energy and centrality dependence of hadron production
from $Au+Au$ collisions.
Results from $pA$ collisions at RHIC could also help to understand some
controversial and puzzling
results from ion-ion collisions, for example the large
azimuthal asymmetry at high $p_\perp$, as
measured by the STAR collaboration.
\end{abstract}

\section{Introduction}\label{intro}
Perturbative (short-distance)
QCD predicts a double-logarithmic distribution of gluons
in a hadron or nucleus at small $x$ (i.e., at rapidities far away from the
valence quarks): $dN_g \sim dk_\perp^2/k_\perp^2\, dx/x$. This result,
however, is derived in the dilute limit where some effects, like gluon
recombination, have been neglected. Therefore, as $k_\perp$ decreases, the
DGLAP approximation should eventually become invalid, and the predicted rise of
the gluon density should flatten out.

Namely, at large distances (though still much
smaller than the confinement scale) the non-linear terms in the Yang-Mills
equations tame the growth of the color charge density in
hadrons~\cite{sat,CGC} as compared to the power-law growth predicted by
the linearized DGLAP theory. The gluon density should saturate
at distances where the color charge density $\rho^a$ per unit
transverse area becomes of order $1/g^2$, where $g$ is the coupling,
and where therefore QCD-evolution becomes genuinely non-perturbative.
Such large occupation numbers are characteristic of a condensate, and so
classical methods may be useful in determining
the color field in a hadron or nucleus at small $x$, i.e.\ at high
density~\cite{sat,CGC} (the ``saturation effect'' is expected
to show up in large nuclei much ``earlier'', at larger $x$, than for
protons). The local color charge density is a
stochastic variable (fluctuating randomly as one moves around in the transverse
plane) and observables eventually have to be averaged over some distribution
of $\rho^a$. This is analogous to spin glasses, and hence the name ``Color
Glass Condensate''~\cite{CGC} for the structure of a nucleus at small $x$.

The Relativistic Heavy-Ion Collider (RHIC) at BNL is dedicated to the study of
high-energy hadronic interactions, from $p+p$ over $p+A$ to $A+B$ collisions.
One of the important issues to settle is whether the above-mentioned gluon
saturation scale $Q_s$ at the values of $x$ that are
relevant at RHIC energy is in fact (substantially) larger than the QCD
confinement scale $\Lambda$. Optimistic estimates~\cite{KL} do indeed yield
$Q_s^2(y=0)\simeq2$~GeV$^2$ for central Au+Au collisions at RHIC, that is
$Q_s^2/\Lambda^2\simeq50$. {\em If} so, then perturbative estimates of
particle production at leading twist~\cite{xnwg} can not be employed to
compute the dominant part of the inelastic cross section
because one needs to resum all powers of $\alpha_s$ times the charge density
squared per unit of transverse area (``higher twists''). In other words,
particle production will be dominated by the decay of the classical non-abelian
field, except at very short distances (truly high $p_t$) where gluon occupation
numbers are small and one enters the region of applicability of perturbation
theory. The initial conditions for the subsequent evolution, which possibly
proceeds through a so-called Quark-Gluon Plasma, then have to be
obtained from the Color Glass Condensate model~\cite{KrV}.
On the other hand, if it turns
out that at RHIC energy $Q_s$ is {\em not} substantially larger than $\Lambda$,
as some conservative estimates indicate~\cite{gymcl}, then this means that
leading-twist perturbation theory is safe at RHIC, and that it can be
employed to describe the bulk of particle production~\cite{footn1}.
To understand particle multiplicities and
energy densities in the central region one does not need to consider
the soft classical field then~\cite{footn2}.

Naturally, as RHIC started operation with collisions of gold ions, theoretical
studies focused on the energy and centrality dependence of particle
production from the CGC model versus that from leading twist perturbation
theory with fixed infrared cutoff $p_0\simeq 2$~GeV~\cite{KL,wg_b,EKRT}.
Within the CGC model, a rather mild increase of $\d N_{ch}/\d \eta$ both
with centrality (or the number of participants) and with energy is expected.
In contrast,
phase space for particle production at $p_\perp>p_0$ rapidly opens up in
perturbation theory, and so the growth is faster. The energy dependence of
particle production at RHIC turns out to be quite weak~\cite{E_dNdy},
approximately consistent with saturation models~\cite{KL,EKRT}. However,
to my opinion it is hard for both saturation models and soft-hard
two-component models to make absolute predictions on
the 10-20\% level (at present),
which would be required to falsify one or the other model
using just the limited energy range of RHIC. Similar concerns apply to the
centrality dependence of particle production~\cite{phenix}. This observable is
plagued in particular by the uncertainties in determining experimentally
the number of
participants for large impact parameter collisions. In that respect, it would
be cleaner to determine $\d N_{ch}/\d\eta/N_{part}$ by using only central
collisions but varying the mass number of the colliding nuclei. Of course,
this will not be available within the near future, although it is doable in
principle at RHIC, which can collide essentially any ions from $p$ to $Au$.
 
Is it possible then to verify or falsify the gluon saturation in nuclei at
RHIC energies~? I think a good opportunity for that exists in $pA$ since
protons are likely quite dilute at RHIC energy, while saturation may occur
for large nuclei~\cite{kov_muel,ad_larry}.
In that case, distinct signatures emerge in the
transverse momentum and rapidity distributions of secondaries, which are
correlated (and so can be tested experimentally without necessarily relying
on absolute predictions). Also, one might be able to test various scalings of
observables like $\d N/\d y$, $\langle p_\perp\rangle$ and so on.
More detailed observables like the $p_\perp$-broadening of minijets at
forward rapidities~\cite{ad_jamal}, or diffractive single-jet
production~\cite{diffr} may as well provide deeper insight.

\section{Proton-Nucleus Collisions}\label{sect_pA}
$pA$ collisions at high energy are very interesting because one can
actually compute the total multiplicity analytically~\cite{kov_muel,ad_larry},
and also make some
predictions for transverse momentum spectra and the scaling of $\langle
p_\perp\rangle$ with the produced multiplicity $dN/dy$.

The key point is that the color charge density per unit transverse area is
much smaller in a proton than in a big nucleus. Therefore, the saturation
momentum of the proton is much smaller than that of the nucleus, $Q_s^{(1)}
\ll Q_s^{(2)}$. At $p_\perp
\gg Q_s^{(2)}$ the fields of both the proton and of the
nucleus are weak, and so one expects that ordinary perturbation theory is
applicable. In particular, to compute single-inclusive gluon production at
high $p_\perp$ the well-known expressions from collinear factorization (with
DGLAP evolution) can be employed.

On the other hand, the inelastic cross section (particle production) is
dominated by the kinematic region $Q_s^{(1)}<p_\perp
<Q_s^{(2)}$. In that regime,
the field of the proton is still weak, and so can be treated within
perturbation theory; but the field of the nucleus is in the non-linear regime,
and one must resum interactions of the radiated gluons with the background
field of the nucleus to all orders. A simple way to understand why is to
realize that while each additional gluon ``line'' from the produced gluon to
the nuclear field comes with an additional power of the coupling $\alpha_s$,
the occupation numbers of the nuclear field at (or below) the scale
$Q_s^{(2)}$ are of order $1/\alpha_s$, and so additional ``rescatterings'' are
{\em not} suppressed by powers of the coupling constant.

The calculation of gluon production essentially consists then of solving the
Yang-Mills equations in the forward light-cone to all orders in the background
field of the nucleus, but to leading order in the field of the weak proton.
From the Fourier transform of the field at asymptotic times one reads off the
amplitudes of the modes, which are then squared and averaged over the
color charge distributions in the proton and
in the nucleus, using a Gaussian weight~\cite{sat}. Finally, this leads to the
gluon distribution~\cite{ad_larry}
\be
\frac{\d N}{\d^2b \d^2k_\perp\d y} =
  2g^2\chi_1(y) \frac{N_c^2-1}{(2\pi)^4} 
 \int \d^2u_\perp
\int\limits_{\Lambda_{\rm QCD}^2} \frac{\d^2p_\perp}{2\pi}
\frac{e^{i (p_\perp-k_\perp)\cdot u_\perp}}{p_\perp^2}
 \exp\left\{g^4N_c\gamma(u_\perp)\chi_2(y)
\right\}~, \label{main_result}
\ee
with
\be
\gamma(x_\perp) = \int\limits_{\Lambda_{\rm QCD}^2}\frac{\d^2q_\perp}{(2\pi)^2}
\frac{e^{i q_\perp\cdot x_\perp}}{q_\perp^4}
\ee
the gluon propagator. $\chi_i(y)$ denote the effective densities of color
charge as ``seen'' by the gluon produced at rapidity $y$. It is the
{\em integrated} charge density of either one of the sources, from its
beam rapidity down to $y$. Note that~(\ref{main_result}) is linear in
$\chi_1$ but resums all orders in $\chi_2$. A numerical evaluation of the
above result has not yet been performed, mainly because one needs to determine
the dependence of $\chi_2$ on rapidity~\cite{KL}. One can nevertheless gain
some insight by considering two limits. Namely, at asymptotically
large $k_\perp$, the exponential can be expanded to first
non-trivial order, leading to the well-known pQCD result
\be
 \frac{\d N}{\d^2b \d^2k_\perp\d y} =
 \frac{4g^6N_c(N_c^2-1)}{(2\pi)^4} \frac{\chi_1(y)\chi_2(y)}{k_\perp^4}
\log\frac{k_\perp}{Q_s^{(2)}}~, \label{pert_kt4}
\ee
where we introduced the shorthand notation
$Q_s^{(1,2)}(y)\equiv\sqrt{g^4N_c\chi_{1,2}(y)/8\pi}$~\cite{footn4}.
Thus, one recovers the standard perturbative $\sim \alpha_s^3/k_\perp^4$
behavior at very high $k_\perp$, with a logarithmic correction analogous to
DGLAP evolution. Note that $\chi_1$, $\chi_2$ scale as $A_1^{1/3}$ and
$A_2^{1/3}$~\cite{gymcl}, respectively~\cite{footn3},
while the integral over $\d^2b$ gives
a factor of $\pi R^2_2\propto A_2^{2/3}$. Therefore, in this kinematic
region $\d N/\d^2k_\perp\d y$ scales like $A_1^{1/3} A_2$, up to
logarithmic corrections. This holds also
for the integrated distribution $\d N(k_\perp>p_0)/\d y$ above some
fixed $A_2$-independent scale $p_0$, as in the soft-hard two-component model.
On the other hand, when integrating over $k_\perp$ from
$Q_s^{(2)}$ to infinity, the contribution from large $k_\perp$ to
the rapidity density is
\be \label{dNdy_pert}
\frac{\d N}{\d^2b \d y} = \frac{g^2 (N_c^2-1)}{\pi^2}
 \chi_1(y)~.
\ee
Again, the integral over $\d^2b$ gives a factor $\pi R_2^2\sim
A_2^{2/3}$, and so $\d N/\d y$ scales like $A_1^{1/3} A_2^{2/3}$.

\begin{figure}[htp]
\centerline{\hbox{\epsfig{figure=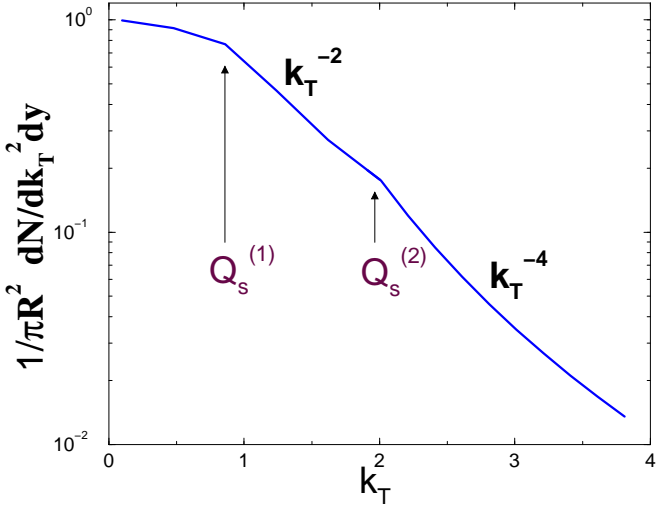,width=8.5cm}}}
\caption{Schematic $k_\perp$ distribution for gluons produced in
high-energy $A_1+A_2$
collisions at rapidity $y$ such that $Q_s^{(1)}(y)\ll Q_s^{(2)}(y)$.
In the perturbative regime, $\d N/\d k_\perp^2\d y\sim1/k_\perp^4$.
In between the saturation scales for the two sources,
$\d N/\d k_\perp^2\d y\sim1/k_\perp^2$.}
\label{figkt}
\end{figure}

On the other hand, when $Q_s^{(1)}(y)\ll k_\perp\ll Q_s^{(2)}(y)$, one
finds that
\be
\frac{\d N}{\d^2b \d^2 k_\perp \d y}
\simeq 2 g^2\chi_1(y)\frac{N_c^2-1}{(2\pi)^3}
 \frac{1}{k_\perp^2}\log\frac{k_\perp}{Q_s^{(1)}(y)}~.
 \label{sat_kt2}
\ee
This form $\sim \alpha_s\chi_1/k_\perp^2$ is to be compared with that from
eq.~(\ref{pert_kt4}), $\sim \alpha_s^3\chi_1\chi_2/k_\perp^4$, valid at high
$k_\perp$.
A schematic distribution
in transverse momentum is shown in Fig.~\ref{figkt}.

Although the figure does not provide a quantitative
prediction for experiments at RHIC, one should notice non-trivial features.
It is predicted that when strong-field effects set in around $Q_s^{(2)}$,
the $k_\perp$ distribution of secondaries flattens. Not only does the scaling
with $A_2$ differ in the two regimes, as mentioned above, but more importantly
the ``turnover'' point $Q_s^{(2)}(y)$ is a function of rapidity~!
Experimentally, one can thus take the rapidity dependence of the saturation
momentum from a parametrization of HERA data~\cite{GBW}, which also seems to
fit the observed rapidity dependence of $\d N/\d y$ from $Au+Au$ at
RHIC~\cite{KL},
and test whether the turnover in the {\em transverse momentum distribution}
from fig.~\ref{figkt} moves as one changes the {\em rapidity}
in a way consistent with those parametrizations.

From~(\ref{sat_kt2}), the $k_\perp$-integrated multiplicity in the
nonperturbative regime $Q_s^{(1)}(y) \le k_\perp\le Q_s^{(2)}(y)$ is
\be \label{dNdy_sat}
\frac{\d N}{\d^2b \d y} =
g^2\chi_1(y)\frac{N_c^2-1}{(2\pi)^2}
\log^2\frac{Q_s^{(2)}(y)}{Q_s^{(1)}(y)}~.
\ee
Thus, at fixed impact parameter, the multiplicity scales as
$A_1^{1/3}$, up to the square of a logarithm of $(A_2/A_1)^{1/3}$.

The average transverse momentum in the saturation regime is given by
\be
\langle k_\perp\rangle = 2\; Q_s^{(2)}\;
\frac{\xi-1-\log\xi}{\log^2\xi}~,   \label{mkperp_sat1}
\ee
where $\xi(y)=Q_s^{(1)}(y)/Q_s^{(2)}(y)$. 
From dimensional considerations, it has been suggested~\cite{juergen} that in
symmetric $A+A$ collisions, and at
central rapidity, $\langle k_\perp\rangle^2$ scales with the multiplicity per
unit of transverse area and of rapidity, $
\langle k_\perp\rangle^2 \propto {\d N}/{\d^2b\d y}$.
A similar scaling relation can be derived from
eqs.~(\ref{dNdy_sat},\ref{mkperp_sat1}) for the asymmetric case,
\be \label{mkperp}
\langle k_\perp\rangle^2 \propto 
\frac{\d N}{\d^2b\d y} \; \frac{g^2}{\xi^2} \;
\frac{\left(\xi-1-\log\xi\right)^2}{\log^6\xi}~.
\ee
Thus, $\langle k_\perp\rangle^2$ is
proportional to the multiplicity per unit of rapidity and
transverse area, times a function of the ratio of the saturation momenta.
If source one is very much weaker than source two, i.e.\ in the limit
$|\log\xi|\gg1-\xi$, the third factor on the
right-hand-side of~(\ref{mkperp}) depends on $\log\xi$ only.
Neglecting that dependence, and assuming as before that $\chi_{1,2}$ are
proportional to $A_{1,2}^{1/3}$, one has the approximate scaling relation
\be \label{simple_scaling}
\langle k_\perp\rangle^2 \propto \left(\frac{A_2}{A_1}\right)^{1/3}
\frac{\d N}{\d^2b\d y}\propto
\left(\frac{1}{A_1 A_2}\right)^{1/3} \frac{\d N}{\d y}~.
\ee
In practice though one expects significant corrections to the simple scaling
relation~(\ref{simple_scaling}), as given by eq.~(\ref{mkperp}). $pA$
collisions provide a natural testing ground for the scaling of
$\langle p_\perp\rangle$ with $\d N/\d y$ since distortions of the above
scaling relations from final-state interactions
(``hydrodynamic expansion'') should not be a major issue.

\begin{table}[hbt] 
\caption[]{Scaling of various quantities in the perturbative
and saturation kinematic regimes.}\label{tab_sat}
\begin{center}
\begin{tabular}{lll}
\hline\\[-10pt]
  & sat. & pert. \\
\hline\\[-10pt]
$\d N/\d^2b\d^2k_\perp\d y$ & $A_1^{1/3}$ & $A_1^{1/3}A_2^{1/3}$ \\
$\langle k_\perp\rangle^2$ & $\left( A_1A_2\right)^{-1/3} \d N/\d y$ & 
                           1 (for fixed $p_0$ scale) \\ 
$\frac{\d}{\d y} \log \frac{\d N}{\d^2k_\perp\d y}$ & $\frac{\d}{\d y} \log 
  Q_s^{(1)}(y)$ & $\frac{\d}{\d y} \left[ \log Q_s^{(1)}(y)+\log Q_s^{(2)}(y)
   \right] $ \\ 
\hline 
\end{tabular}
\end{center}
\end{table}
For rapidities far from the fragmentation region of the large
nucleus, and for $Q_s^{(1)}(y) \ll k_\perp\ll Q_s^{(2)}(y)$,
$\d N/\d^2k_\perp\d y$ varies with rapidity like
\be
\frac{\d}{\d y}\frac{\d N}{\d^2k_\perp\d y} \propto g^2 \chi_1'(y)~.
\ee
Thus, an experimental measure for the evolution of the CGC density
parameter in rapidity is
\be \label{RGchi1}
\frac{\d\log \d N/\d^2k_\perp\d y}{\d y} =
\frac{\d\log\chi_1(y)}{\d y}~.
\ee
Now consider the case of high $k_\perp \gg Q_s^{(2)}(y)$ described
by eq.~(\ref{pert_kt4}). In that regime the rapidity distribution is
proportional to $\chi_1(y)\chi_2(y)$, and so $\d N/\d y\d^2k_\perp$ varies
with rapidity like
\be \label{RGchi12}
\frac{\d\log \d N/\d^2k_\perp\d y}{\d y} = \frac{\d\log\chi_1(y)}{\d y}
  + \frac{\d\log\chi_2(y)}{\d y}~.
\ee
Subtracting~(\ref{RGchi1}) from~(\ref{RGchi12}) provides an experimental
measure for the renormalization-group evolution of $\chi_2(y)$. This does not
only provide an independent verification of the HERA fits~\cite{GBW}, but
also a consistency check on how the turnover point (where the distribution
flattens) from fig.~\ref{figkt} moves as one goes to different rapidity~!
In table~\ref{tab_sat}
I have listed a few scaling relations of various quantities.

\begin{figure}[htp]
\centerline{\hbox{\epsfig{figure=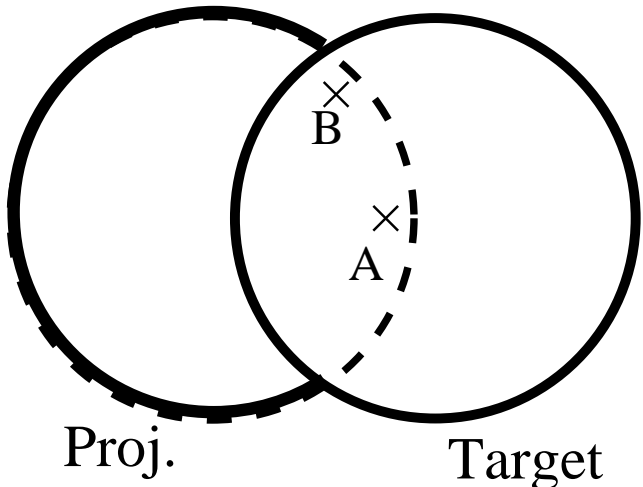,width=8.5cm}}}
\caption{Illustration of a collision of equal-size nuclei at non-zero
impact parameter. Near point B, close to the edges of {\em both} nuclei,
the local densities of partons in the initial state are small; there,
the conditions resemble those for $pp$ collisions. Near point A, we are
encountering a $pA$-like situation: the local parton density near the edge of
the projectile is small but it is large in the target (close to the center).}
\label{fig_ellip}
\end{figure}

In fact, the results obtained above may even be relevant for collisions of
equal-size nuclei, like $Au+Au$. Namely, if we
consider a collision at non-zero impact parameter, as shown in
fig.~\ref{fig_ellip}, it is easy to realize that near point A in that figure
the situation is precisely the one considered above for $pA$ collisions:
a low-density projectile colliding with a high-density target.
From the discussion above, the $k_\perp$ distribution of gluons produced
in the impact is harder near point A than near point B, due to saturation
effects in the target. Now, if those gluons suffer from radiative energy
loss as they interact with each other, it is clear that a detector positioned
at the top of the page will detect less momentum flowing into it than a
detector positioned at the right of the page. Thus, radiative energy loss
will generate an azimuthal asymmetry
\be
v_2(p_\perp) = \frac{\langle p_\rightarrow^2 - p_\uparrow^2\rangle}
{p_\perp^2}
\ee
at transverse momenta on the order of a few GeV~\cite{v2_eloss}. The point
here is that $v_2$ will be {\em larger} if the $p_\perp$-distribution of
produced gluons is harder at A than at B as compared to the case when
$\d N/\d^2 p_\perp$ is the same at A and B. This might be important for
understanding the large values of $v_2$ measured by STAR at RHIC at
$p_\perp\simeq 2-6$~GeV~\cite{STAR_v2},
as it has been pointed out that large energy loss
by itself (assuming the same $p_\perp$-distribution throughout the overlap
zone in fig.~\ref{fig_ellip}) does not lead to large enough
$v_2$~\cite{v2max}, in particular for semi-central events (10\% centrality
class). Quantitative computations of $v_2$ from the gluon saturation model
are presently under way~\cite{v2_sat}.

So far, all attempts to understand the large $v_2$ at high-$p_\perp$ invoke
radiative parton energy loss~\cite{E-loss}
\be
-\Delta E \simeq \frac{3\alpha_s}{\pi} E_{cr} \log \frac{2E}{L m_\ell^2}~.
\ee
Here, $L$ is the typical length of the medium traversed by the jet, $m_\ell$
is the IR cutoff provided by the mass of static electric gluons, $E>E_{cr}$
is the jet energy, and $E_{cr}=m^2_\ell L^2/\lambda$.
$\Delta E/E$ depends on the density of the medium via the mean-free path
$\lambda$ of the radiation. Going down in
$\sqrt{s}$, presumably this should reduce $v_2$ at high $p_\perp$ (in the sense
that the leading-twist prediction $v_2=0$ should be approached faster).
Is this the case~?

Finally, if indeed $v_2$ at $p_\perp=2-6$~GeV is generated by final-state
interactions of the minijets, it might be sensitive to nearly
critical scattering at $T_c$~\cite{ad_rob}, which would evidently be very
exciting. QCD with 3 colors is not an ideal gas near $T_c$; rather, the
pressure $p(T)$ is significantly smaller than the ideal-gas pressure
$p_{id}(T)$. The transition from confinement to deconfinement appears to be
{\em nearly a second order} transition~\cite{lat_2ndO}, in which case 
$(m_\ell/T)^2 \sim \sqrt{p/p_{id}}$ drops rapidly as $T_c$ is approached from
above~\cite{ad_rob}. This behavior is quite different from a strong first-order
transition, and can enhance energy loss (and $v_2$) near $T_c$.

\section{Conclusions}\label{concl}
$Au+Au$ collisions from RHIC {\em may} have offered a first glance at
gluon saturation effects in large nuclei: experimentally, the energy and
centrality dependence of particle production is quite weak.
However, in my opinion it is too early yet to draw definite conclusions
from the $Au$-runs at $\sqrt{s}=130$~AGeV and 200~AGeV. Collisions at lower
energy and in particular with different projectile/target combinations should
be analyzed in the future. Collisions of small projectiles on large
targets, for which I use the generic acronym ``pA'', could reveal distinct
scaling laws predicted by the Color Glass Condensate model.

\section*{Acknowledgement(s)}
Support from DOE Grant DE-AC02-98CH10886 is gratefully acknowledged.
I would like to thank M.~Gyulassy, J.~Jalilian-Marian, D.~Kharzeev,
L.~McLerran, D.~Teaney, and R.~Venugopalan for many stimulating discussions.
Many thanks also to the organizers of this very pleasant workshop for
inviting me to present this work. 

\vfill\eject

\begin{thebibliography}{99}
\bibitem{footn2}
Moreover, the classical approach is of
little use then because if $Q_s\simeq\Lambda$, weak-coupling methods can not
be applied and so hardly anything reliable can actually be computed. Note that
this is a {\em physical} requirement for the classical effective theory to
hold, not a technical one, and so is equally mandatory for numerical
lattice implementations of the theory~\cite{KrV}.
\bibitem{footn4}
This $Q_s$ is larger than that from~\cite{KL} by a factor of
$\sqrt{2\pi^2N_c/(N_c^2-1)}\simeq2.7$.
\bibitem{footn3}
We have introduced the mass number $A_1$ of the projectile, here, which does
not need to be a proton. All that matters is that the projectile is much
smaller than the target, such that the ratio of their saturation scales is
far from unity. One can equally well consider $d+Au$ collisions, such that
$A_1=2$ etc.
\end{thebibliography}
\end{document}